\documentclass[aps,
showpacs,tightenlines]{revtex4}
\usepackage{graphics}
\begin{document}
\bibliographystyle{apsrev}

\newcommand{\half}{\frac{1}{2}}
\newcommand{\D}{\mbox{D}}
\newcommand{\curl}{\mbox{curl}\,}
\newcommand{\ep}{\varepsilon}
\newcommand{\lleq}{\lower0.9ex\hbox{ $\buildrel < \over \sim$} ~}
\newcommand{\ggeq}{\lower0.9ex\hbox{ $\buildrel > \over \sim$} ~}
\newcommand{\tr}{{\rm tr}\, }

\newcommand{\be}{\begin{equation}}
\newcommand{\ee}{\end{equation}}
\newcommand{\bea}{\begin{eqnarray}}
\newcommand{\eea}{\end{eqnarray}}
\newcommand{\beaa}{\begin{eqnarray*}}
\newcommand{\eeaa}{\end{eqnarray*}}
\newcommand{\Lhat}{\widehat{\mathcal{L}}}
\newcommand{\nn}{\nonumber \\}
\newcommand{\e}{{\rm e}}


\title{Dark energy cosmology from higher-order string-inspired gravity
and its reconstruction}
\author{Shin'ichi Nojiri}
\email{nojiri@phys.nagoya-u.ac.jp, snojiri@yukawa.kyoto-u.ac.jp}
\affiliation{Department of Physics, Nagoya University, Nagoya 464-8602. Japan}
\author{Sergei D. Odintsov\footnote{also at Lab. Fundam. Study, Tomsk State
Pedagogical University, Tomsk}}
\email{odintsov@ieec.uab.es}
\affiliation{Instituci\`{o} Catalana de Recerca i Estudis Avan\c{c}ats (ICREA)
and Institut de Ciencies de l'Espai (IEEC-CSIC),
Campus UAB, Facultat de Ciencies, Torre C5-Par-2a pl, E-08193 Bellaterra
(Barcelona), Spain}
\author{M. Sami}
\email{sami@iucaa.ernet.in}
\affiliation{Centre for Theoretical Physics, Jamia Millia Islamia, New Delhi
and Department of Physics, Jamia Millia Islamia, New Delhi, India}

\begin{abstract}
In this paper we investigate the cosmological effects of modified
gravity with string curvature corrections added to Einstein-Hilbert
action in the presence of a dynamically evolving scalar field
coupled to Riemann invariants. The scenario exhibits several
features of cosmological interest for late universe. It is shown
that higher order stringy corrections can lead to a class of dark
energy models consistent with recent observations. The model can
give rise to quintessence 
without recourse to scalar field potential. The detailed
treatment of reconstruction program for general scalar-Gauss-Bonnet
gravity is presented for any given cosmology. The explicit examples
of reconstructed scalar potentials are given for accelerated
(quintessence, cosmological constant or phantom) universe. Finally,
the relation with modified $F(G)$ gravity is established on
classical level and is extended to include third order terms on
curvature.
\end{abstract}


\maketitle
\section{Introduction}
One of the most remarkable discoveries of our time is related to the late time
acceleration of our universe which is supported by observations of high
redshift type Ia supernovae treated as standardized candles and, more
indirectly,
by observations of the cosmic microwave background and galaxy clustering.
The criticality of universe supported by CMB observations fixes the total
energy budget of universe. The study of large scale structure reveals that
nearly $30$ percent of the total cosmic budget is
contributed by dark matter. Then there is a deficit of
almost $70$ percent; in the standard paradigm, the missing component is an
exotic form
of energy with large negative pressure dubbed
{\it dark energy} \cite{d1,d2,d,CST,review,reviewI,reviewIV}.
The recent observations on baryon
oscillations provide yet another independent support to dark energy
hypothesis\cite{Eis}.

The dynamics of our universe is described by the Einstein equations in which
the contribution of energy content of universe is
represented by energy momentum tensor appearing on RHS of these equations. The
LHS represents pure geometry
given by the curvature of space-time. Gravitational equations in their original
form
with energy-momentum
tensor of normal matter can not lead to acceleration. There are then two ways
to obtain accelerated expansion, either by supplementing energy-momentum tensor
by dark energy component or by modifying the geometry
itself.

Dark energy problem is one of the most important problems of modern
cosmology and despite of the number of efforts (for a recent review,
see \cite{CST,d}), there is no consistent theory which may
successfully describe the late-time acceleration of the universe.
General Relativity with cosmological constant does not solve the
problem because such theory is in conflict with radiation/matter
domination eras. An alternative approach to dark energy is related
to modified theory of gravity (for a review, see \cite{rev3}) in
which dark energy emerges from the modification of geometry of our
universe. In this approach, there appears quite an interesting
possibility to mimic dark energy cosmology by string theory. It was
suggested in refs.\cite{Nojiri:2005vv,Sami:2005zc,Mota} that dark energy
cosmology may result from string-inspired gravity. In fact,
scalar-Gauss-Bonnet gravity  from bosonic or Type II strings was
studied in the late universe \cite{Nojiri:2005vv,Sami:2005zc} (for
review of the applications of such theory in the early universe ,
see \cite{Calcagni:2005im}). It is also interesting such theories
may solve the initial singularity problem of the standard big-bang
model(see \cite{ART} and refs. therein). Moreover, the easy account
of next order (third order, Lovelock term) is also possible in this
approach (for recent discussion of such gravity, see \cite{DM}).

In this paper we examine string-inspired gravity with third order
curvature corrections (scalar-Gauss-Bonnet term and scalar-Euler term) and
explore the cosmological dynamics of the
system attributing special attention to dark energy (non-phantom/phantom)
solutions. We confront our result with the recent observations.
We also outline the general program
of  reconstruction of  scalar-Gauss-Bonnet gravity
   for any {\it a priori} given cosmology following the method \cite{e}
developed in the scalar-tensor theory.

The paper is organized as follows. In section two, we consider the
cosmological dynamics in the presence of string curvature
corrections to Einstein-Hilbert action. We analyze cosmological
solutions in the FRW background;
  special attention is paid at dark energy which naturally arises in the model
  thanks to higher order curvature terms induced by string corrections. Brief
discussion on the comparison of theoretical results with recent
observations is included. The stability of dark energy solution is
investigated in detail.

Section three is devoted to the study of late-time cosmology for
scalar-Gauss-Bonnet gravity motivated by string theory but with the
arbitrary scalar potentials. It is explicitly shown how such theory
(actually, its potentials) may be reconstructed for any given
cosmology. Several explicit examples of dark energy cosmology with
transition from deceleration to acceleration and (or) cosmic
speed-up  (quintessence, phantom or de Sitter) phase or with
oscillating (currently accelerating) behavior of scale factor are
given. The corresponding scalar potentials are reconstructed. It is
shown how such theory may be transformed to modified Gauss-Bonnet
gravity which turns out to be just specific parametrization of
scalar-Gauss-Bonnet gravity on classical level. Finally, it is shown
how to include third order  curvature terms in the above
construction. Summary and outlook are given in the last section.

\section{Dark energy from higher order string curvature corrections}
In this section we shall consider higher order curvature corrections to
Einstein-Hilbert action.
To avoid technical complications we restrict the discussion up to third order
Riemann invariants
coupled to a dynamical field $\phi$. The cosmological dynamics of the system
will be developed in detail and general features of the solutions will be
discussed. It is really
interesting that the model can account for recent observations on dark energy.

\subsection{General action}

We begin from the following action
\be
\label{act}
{\cal S}=\int d^4x \sqrt{-g}\left[
\frac{R}{2\kappa^2}-\frac12\omega(\phi)g^{\mu \nu} \partial_\mu \phi
\partial_\nu \phi-V(\phi)+{\cal  L}_c+{\cal L}_m\right]\ ,
\ee
where $\phi$ is
a scalar field which, in particular case, could be a dilaton.
${\cal L}_m$ is the Lagrangian of
perfect fluid with energy density $\rho_m$ and pressure $p_m$.
Note that scalar potential coupled to curvature (non-minimal coupling)
\cite{faraoni} does not appear in string-inspired gravity in the frame under
consideration.

The quantum corrections are encoded in
the term
\be
{\cal L}_c=\xi_1(\phi){\cal L}_c^{(1)}+\xi_2(\phi){\cal L}_c^{(2)}
\ee
where $\xi_1(\phi)$ and $\xi_2(\phi)$  are the couplings of
the field $\phi$ with higher curvature invariants. ${\cal L}_c^{(1)}$ and
${\cal L}_c^{(2)}$ are given by
\begin{eqnarray}
\label{Lc1}
&& {\cal L}_c^{(1)} =R_{\alpha\beta\mu\nu}R^{\alpha\beta\mu\nu}
-4R_{\mu\nu}R^{\mu\nu}
+R^2 \,, \\
&& {\cal L}_c^{(2)}=E_3+R^{\mu\nu}_{\alpha \beta}
R^{\alpha\beta}_{\lambda\rho}
R^{\lambda\rho}_{\mu\nu}
\label{Lc2}
\end{eqnarray}
The third order Euler density $E_3$ is proportional to
\be
\label{E1}
E_3\propto  \epsilon^{\mu\nu\rho\sigma\tau\eta}
\epsilon_{\mu'\nu'\rho'\sigma'\tau'\eta'}
R_{\mu\nu}^{\ \ \mu'\nu'} R_{\rho\sigma}^{\ \ \rho'\sigma'}
R_{\tau\eta}^{\ \ \tau'\eta'}\ .
\ee
Since there does not exist $\epsilon^{\mu\nu\rho\sigma\tau\eta}$ if
the space time dimension $D$ is less than $6$; $E_3$ should
vanish when $D<6$, especially in four dimensions.
By using
\be
\label{E2}
\epsilon^{\mu\nu\rho\sigma\tau\eta}
\epsilon_{\mu'\nu'\rho'\sigma'\tau'\eta'}
= \delta_\mu^{\ \mu'}\delta_\nu^{\ \nu'}\delta_\rho^{\ \rho'}
\delta_\sigma^{\ \sigma'}\delta_\tau^{\ \tau'}\delta_\eta^{\ \eta'}
\pm \left(\mbox{permutations}\right)\ ,
\ee
we can rewrite the expression (\ref{E1}) as
\begin{eqnarray}
E_3 &\propto& 8\left(R^3 - 12 R R_{\mu\nu} R^{\mu\nu}
+ 3 R R_{\mu\nu\rho\sigma}R^{\mu\nu\rho\sigma} + 16 R_\mu^{\ \nu}R_\nu^{\ \rho}
R_\rho^{\ \mu}
+ 24 R_\mu^{\ \nu} R_\rho^{\ \sigma} R_{\nu\sigma}^{\ \ \mu\rho}  \right. \nn
&& \left. - 24 R_\mu^{\ \nu}R_{\nu\rho}^{\ \ \sigma\tau} R_{\sigma\tau}^{\ \
\mu\rho}
+ 2 R_{\mu\nu}^{\ \ \rho\sigma} R_{\rho\sigma}^{\ \ \tau\eta}
R_{\tau\eta}^{\ \ \mu\nu} - 8 R_{\mu\nu}^{\ \ \rho\tau} R_{\rho\sigma}^{\ \
\mu\eta}
R_{\tau\eta}^{\ \ \nu\sigma} \right)\ .
\label{E3}
\end{eqnarray}
We should note in the r.h.s. of (\ref{E2}), there appears $6!=720$ terms,
which correspond to the sum of the absolute values of the coefficients in each
term
in the RHS of (\ref{E3})
\be
\label{E4}
8\left(1+12+3+16+24+24+2+8\right)=720\ .
\ee
In what follows we shall be interested in the cosmological applications of
modified equations
of motion and  thus assume a flat Friedmann-Robertson-Walker (FRW) metric
\be\label{FRW}
ds^2=-N^2(t) dt^2+ a^2(t)\sum_{i=1}^d (dx^i)^2,
\ee
where $N(t)$ is the lapse function.
With the metric (\ref{FRW}), the Riemann invariants read
\be
\label{Lc1frw}
{\cal L}_c^{(1)}=24 H^2\left(\frac{\dot{H}+H^2}{N^4}-\frac{\dot{N}}{N^5} H
\right)\,, \quad
{\cal L}_c^{(2)}=\frac{24}{N^6}(H^6+I^3)-\frac{72\dot{N}}{N^7}HI^2
\label{Lc2frw}
\ee
where $I=\dot{H}+H^2$ and $H=\dot{a}/a$. It is straightforward though
   cumbersome to verify explicitly that third order
Euler density $E_3$ is identically zero in the FRW background.
The non-vanishing contribution in Eq.(\ref{Lc2frw}) comes from the second term
in (\ref{Lc2}).
To enforce the check in a particularly case, we consider $D$ dimensional
de-Sitter space, where
Riemann curvature is given by
\be
\label{E5}
R_{\mu\nu}^{\ \ \rho\sigma}=H_0\left(\delta_\mu^{\ \rho} \delta_\nu^{\ \sigma}
   - \delta_\mu^{\ \sigma} \delta_\nu^{\ \rho}\right)\ .
\ee
Here $H_0$ is a constant corresponding to the Hubble rate. In the de-Sitter
background we have
\be
\label{E6}
E_3 \propto  D(D-1)(D-2)(D-3)(D-4)(D-5)\ ,
\ee
   which is obviously zero in case of $D<6$.
For simplicity we shall limit the discussion to a
homogeneous scalar field $\phi(t)$.
Then the spatial volume can be integrated out from
the measure in equation (\ref{act}), which we rewrite as
\be
{\cal S}=\int{ dt Na^3
\left[\frac{R}{2 \kappa^2}+{\cal L}_c+{\cal L}_{\phi}+{\cal L}_m\right]}.
\label{Naction}
\ee
where ${\cal L_{\phi}}=-\frac12\omega(\phi)(\nabla\phi)^2-V(\phi)$.
Varying the action (\ref{Naction}) with respect to the lapse function $N$
we obtain \cite{Sami:2005zc}
\begin{equation}
\frac{3H^2}{\kappa^2}=\rho_m+\rho_{\phi}+\rho_c
\label{qhubble}
\end{equation}
where
\begin{eqnarray}
\rho_{\phi}=\frac{1}{2}\omega \dot{\phi}^2+V(\phi)
\end{eqnarray}
In Eq.(\ref{qhubble}), the energy density $\rho_c$ is induced by quantum
corrections and is given by
the following expression
\be
\rho_c=\left.\left(3H\frac{\partial {\cal L}_c}{\partial
\dot{N}}+\frac{d}{dt}\frac{\partial {\cal L}_c}{\partial
\dot{N}}-\frac{\partial {\cal L}_c}{\partial
N}-{\cal L}_c\right)\right|_{N=1}
\ee
It would be convenient to rewrite  $\rho_c$ as
\be
\rho_c=\xi_1(\phi) \rho_c^{(1)}+\xi_2(\phi)\rho_c^{(2)}
\ee
Using Eqs.(\ref{Lc1frw}) $\&$ (\ref{Lc2frw}) we obtain the expressions
of  $\rho_c^{(1)}$ and $\rho_c^{(2)}$
\begin{eqnarray}
&&\rho_c^{(1)}=-24 H^3\Xi_1 \, \\
&& \rho_c^{(2)}=-72 H I^2 \Xi_2-72\left(\dot{H}I^2+2 IH\dot{I}\right)
-216H^2 I^2+120\left(H^6+I^3 \right)
\end{eqnarray}
where $\Xi_1={\dot{\xi}_1}/{\xi}_1$ and $\Xi_2={\dot{\xi}_2}/{\xi}_2$
It is interesting to note that the contribution of Gauss-Bonnet term (described
by Eq.(\ref{Lc1frw})) cancels in equations of motion
for fixed $\phi$ as it should be; it contributes for dynamically
evolving scalar field only. In case of the third order curvature corrections,
the Euler density is identically zero and hence it does not contribute 
to the equation of motion in general.
Secondly, ${\cal L}_c^{(2)}$ contributes for fixed field as well as for
dynamically evolving
$\phi$. It contains corrections of third order in curvature beyond the Euler
density.

We should note that such higher-derivative terms in string-inspired
gravity may lead to ghosts and related instabilities (for recent
discussion in scalar-Gauss-Bonnet  gravity, see \cite{ghost}).
However, the ghost spectrum of such (quantum ) gravity (for the
review, see\cite{book}) is more relevant at the early universe where
curvature is strong, but less relevant at late universe. Moreover,
in accordance with modified gravity approach, the emerging theory is
purely classical, effective theory which comes from some unknown
gravity which has different faces at different epochs. (Actually, it
could be that our universe currently enters to instable phase). For
instance, in near future the currently sub-leading terms may
dominate in the modified gravity action
  which
then has totally different form! Hence, this is that (unknown)
gravity, and not its classical limit given by Eq.(\ref{act})
relevant during specific epoch, whose spectrum should be studied.
The point is best illustrated by the example of Fermi theory of weak
interactions whose quantization runs into well known problems.
Finally, on the phenomenological grounds, it is really interesting
to include higher order terms. At present the situation is
remarkably tolerant in cosmology, many exotic constructions attract
attention provided they can lead to a viable model of dark energy.

The equation of motion for the field $\phi$ reads from (\ref{Naction})
\be
\omega(\ddot{\phi}+3H\dot{\phi})+V'-\xi_1'{\cal L}_c^{(1)} - \xi_2'{\cal
L}_c^{(2)}
+\dot{\omega}\dot{\phi}-\omega'\frac{\dot{\phi}^2}{2}=0
\label{phieq}
\ee
In addition we have standard continuity equation for the barotropic background
fluid with
energy density $\rho_m$ and pressure $p_m$
\be
\dot{\rho_m}+3H(\rho_m+p_m)=0
\label{conteq}
\ee

Equations (\ref{qhubble}), (\ref{phieq}), and (\ref{conteq}) are the basic
equations for our system under consideration.

Let us note that in the string theory context with the dilaton field $\phi$ we
have
\be
V(\phi)=0,~~\xi_1=c_1\alpha^\prime \e^{ 2 \phi/\phi_0},~
\xi_2=c_2 \alpha^{\prime 2} \e^{4 \phi/\phi_0}
\ee
where $(c_{1}, c_{2})=(0, 0, 1/8), (1/8, 0, 1/8)$
for Type II, Heterotic, and Bosonic strings, respectively.

\subsection{Fixed field case: general features of solutions.}

We now look for de-Sitter solutions in case of $\phi=constant$ and $\rho_m=0$.
In this case the
modified Hubble Eqs.(\ref{qhubble}) gives rise to de-Sitter solution
\begin{equation}
3 =24 \xi_2 H^4~~or~~H=\left(\frac{1}{8\xi_2}\right)^{1/4}
\end{equation}
where $\xi_2=\frac{1}{8}exp(-4\phi/\phi_0)$ for type II and Bosonic
strings. Normalizing $\xi_2$ to one, we find that $H= 0.6$ (we have
set $\kappa^2=1$ for convenience). Below we shall discuss the
stability of the solution. There exists no de-Sitter solution for
Heterotic case. Actually, de-Sitter solutions were investigated in
the similar background in Ref.\cite{Sami:2005zc} where higher order
curvature corrections up to order four were included. Since, here we
confine ourselves up to the third order and the fourth order terms
are excluded from the expression of $\rho_c$; these terms come with
different signs. Thus it becomes necessary to check whether or not
the stability property of de-Sitter solutions is preserved order by
order.

We further note that the modified Hubble Eqs.(\ref{qhubble}) admits the
following solution in
the high curvature regime in presence of the barotropic fluid with equation of
state parameter $w$
\be
a(t) =a_0 t^{h_0} \,, \quad \mbox{or} \quad a(t) =a_0(t_s-t)^{h_0}
\ee
where
\begin{eqnarray}
&& h_0=\frac{2}{1+w} \,, \\
&& a_0=\left[\frac{\xi_2}{\rho_0}\left(72(-h_0I_0^2+2I_0h_0 \dot{I}_0)+
216h_0^2I_0^2-120(h_0^6+I_0^3)\right)\right]^{-\frac {1}{3(1+w)}}
\end{eqnarray}
We have used $\rho_m=\rho_0 a^{-3(1+w)}$ for the background matter density and
$ I_0=h_0(h_0-1),~~\dot{I_0}=-2h_0(h_0-1)$.
For the effective equation of state dictated by the modified Hubble
Eqs.(\ref{qhubble}) we have
\begin{equation}
w_{\rm eff}=-1-\frac {2}{3} \frac{\dot{H}}{H^2}=-1+\frac{1+w}{3}
\label{weff}
\end{equation}
It is interesting  to note that effective EoS parameter (\ref{weff})
may correspond to inflationary solution
in the presence of
background fluid (radiation/matter). In the
low curvature regime
or at late times $w_{\rm eff}=w$. In the presence of
phantom matter, the effective EoS being less than $-1$
is typical for Big Rip singularity.
It is really not surprising that we
have inflationary solution
at early epochs in the presence of higher order curvature correction to Einstein
Hilbert action; an early example
of this phenomenon is provided by $R^2 $-$ gravity$.
\subsection{Autonomous form of equations of motion}
\begin{figure}
\resizebox{3.0in}{!}{\includegraphics{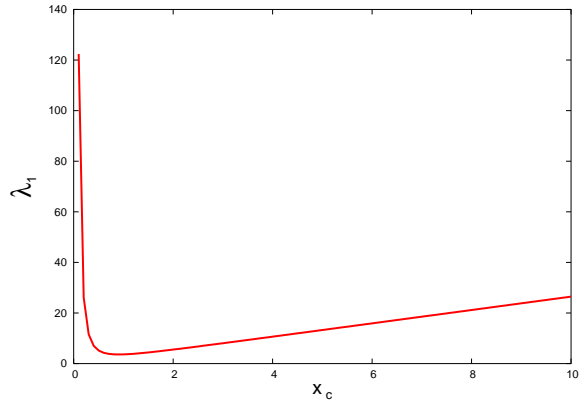}}
\caption{Plot of the first eigenvalue $\lambda_1$ versus the critical point
$x_c$. The
eigenvalue remains positive if the critical point is varied from zero to larger
values.}
\label{eigen}
\end{figure}

Let us now cast the equations of motion in the autonomous form. Introducing
the following notation ($\kappa^2=1$)
\begin{equation}
x= H,~~~y= \dot{H},~~~u= \phi,~~v= \dot{\phi},~~~z= \rho_m
\end{equation}
We shall assume $\omega(\phi)=\nu=const$.
we obtain the system of equations
\begin{eqnarray}
&&\dot{x}=y \,, \nn
&&\dot{y}= \frac{\frac{1}{2}\nu v^2-24\xi_1 \Xi_1 x^3+\xi_2
\left[-72xI^2\Xi_2-72(yI^2+4Iyx^2)-216x^2I^2+120(x^6+I^3)\right]-3x^2}{144
I(x,y) \xi_2 x} +\frac{z}{144I \xi_2 x} \,, \nn
&&\dot{u}=v \,, \quad
\dot{v}=\frac{-3 \nu xv+\xi_1{\cal L}_c^{(1)}+\xi_2 {\cal L}_c^{(2)} }{\nu} \,,
\quad
\dot{z}=-3x(1+w)z
\label{Hddot}
\end{eqnarray}
We shall be first interested in the case of fixed field  
for which we have (assuming $\nu=1$)
\begin{eqnarray}
&&\dot{x}=y \,, \nonumber \\
&&\dot{
y}=\frac{\left[-72(yI(x,y)^2+4I(x,y)yx^2)+216x^2I^2(x,y)+120(x^6+I^3(x,y))-3x^2+z\right]}{144
I(x,y)x}\,, \nonumber\\
&&\dot{z}=-3x(1+w)z
\end{eqnarray}
where
\begin{equation}
I(x,y)=x^2+y,~~~{\cal L}_c^{(1)}=24 x^2(y+x^2),~~~{\cal
L}_c^{(2)}=24(x^6+I^3(x,y)
\end{equation}
In the presence of perfect fluid, the de-Sitter fixed point is characterized by
\begin{eqnarray}
x_c= 0.71,~~~y_c=0,~~~z_c=0
\label{fixedp}
\end{eqnarray}
Perturbing the system around the critical point and keeping the linear terms we
obtain
\begin{eqnarray}
&&\dot{\delta x}=\delta y \,, \nonumber \\
&&\dot{\delta y}=\left(\frac{21}{3}x_c^2+\frac{10}{3 x_c}+\frac{1}{48
x_c^2}\right)\delta x
+\left(\frac{2}{3}x_c+\frac{5}{6 x_c^2}+\frac{1}{48 x_c^3} \right)\delta
y+\frac{1}{144 x_c^3}\delta z \,, \nonumber\\
&&\dot{\delta x}=-3x_c(1+w)\delta z
\end{eqnarray}
Stability of the fixed points depends upon the nature of eigenvalues of
perturbation matrix
\be
\lambda_{1,2}=\frac{1}{2}\left(a_{22}\pm\sqrt{4a_{21}+a_{22}^2}\right) \,,
\quad
\lambda_3=a_{33}=-3x_c(1+w)
\ee
For the fixed point given by (\ref{fixedp}),  $\lambda_{1}$ is positive where
as $\lambda_{2}$ is
negative  making the de-Sitter solution
an unstable node. In fact,  $\lambda_{1}$ remains positive for any $x_c>0$
thereby making the conclusion
independent of the choice of $\xi_2^{(0)}$ (see FIG. \ref{eigen}).
\begin{figure}
\resizebox{3.0in}{!}{\includegraphics{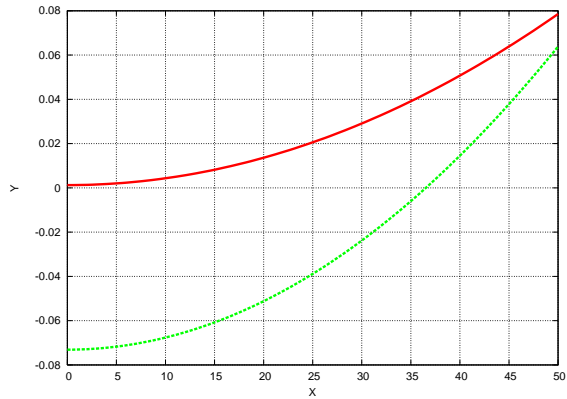}}
\caption{Plot of ${ Y} \equiv (\frac{48 \xi_0^{(1)}}{t_1^2})\times 10^5$ (black
line)
and ${ Y} \equiv (\frac{96 \xi_0^{(2)}}{t_1^4})$ (gray line) versus ${ X} \equiv
\phi_0^2$.
corresponding to $h_0=40$ or $w_{DE}=-0.95$ ($\nu$ is assumed to be one). The common
region corresponding to positive values of the couplings
gives possible models of dark energy induced by higher order curvature
corrections.}
\label{O}
\end{figure}

\begin{figure}
\resizebox{3.0in}{!}{\includegraphics{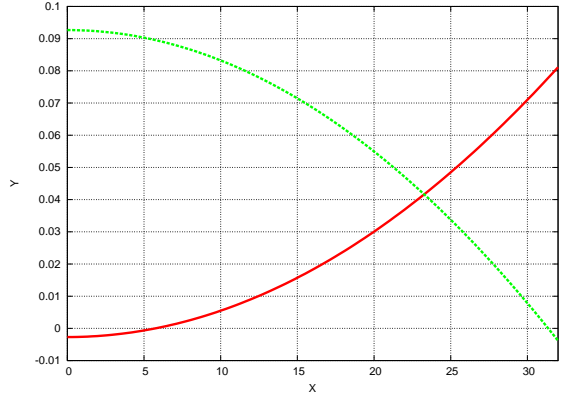}}
\caption{Plot of ${ Y} \equiv (\frac{48 \xi_0^{(1)}}{t_1^2})\times 10^5$ (gray
line)
and ${ Y} \equiv (\frac{96 \xi_0^{(2)}}{t_1^4})$ (black line) versus ${ X} \equiv
\phi_0^2$.
corresponding to $h_0=-33.33$ or $w_{DE} =-1.06$. The region bounded by $6<{ X}<31.5$
corresponds to
possible phantom dark energy models.}
\label{O1}
\end{figure}
\subsection{Dynamically evolving field $\phi$ and dark energy solutions}

In what follows we shall be interested in looking for an
exact solution of equations of motion (\ref{qhubble}) and (\ref{phieq}) which
of
interest to us from the point of view of dark energy in absence of the
background fluid.
In this case let us look for the following solution
\be
H=\frac{h_0}{t},\ \phi=\phi_0\ln\frac{t}{t_1}\ \left(\mbox{when}\
h_0>0\right)\,, \quad
H=\frac{h_0}{t_s-t},\ \phi=\phi_0\ln\frac{t_s-t}{t_1}\ \left(\mbox{when}\
h_0<0\right)\ .
\label{solution}
\ee
Substituting (\ref{solution}) in evolution Eqs. (\ref{qhubble}) and
(\ref{phieq}) yields (we again set $\kappa^2=1$)
\begin{eqnarray}
\label{x1eq1}
&&\nu(1-3h_0)\phi_0^2+\frac{48 \xi_1^{(0)} }{t_1^2} h_0^3(h_0-1)+\frac{96
\xi_2^{(0)}}{t_1^2} (h_0^6+I_0^3)=0 \,, \nn
&&-3h_0^2+\frac{\nu}{2}\phi_0^2-\frac{48
\xi_1^{(0)}}{t_1^2}h_0^3+\frac{\xi_2^{(0)}}{t_1^4}J(h_0)=0
\label{x2eq1}
\end{eqnarray}
where
\begin{eqnarray}
&&J=\frac{1}{96}\left(-288h_0I_0^2-72(-h_0I_0^2+2I_0h_0\dot{I_0})-216h_0I_0^2+120(h_0^6+I_0^3)\right)
\,, \nn
&& I_0=h_0(h_0-1),~~\dot{I_0}=-2h_0(h_0-1)
\end{eqnarray}
Using Eqs.(\ref{x1eq1}), we express the couplings through
$h_0$ and $\phi_0$
\begin{eqnarray}
\label{coupling1}
&&\frac{48 \xi_1^{(0)}}{t_1^2}=\left[\frac {3h_0^2
   -\frac{\nu\phi_0^2}{2}+\nu(3h_0-1)\phi_0^2
}{J(h_0)(h_0-1)+h_0^3(h_0^3+(h_0-1)^3)}\right]\,, \\
&&\frac{96
\xi_2^{(0)}}{t_1^4}=\frac{1}{h_0^3}\left[-3h_0^2+\frac{\nu\phi_0^2}{2}
+\left[\frac
{\left(3h_0^2-\frac{\nu\phi_0^0}{2}+\nu(3h_0-1)\phi_0^2\right)J(h_0)}{J(h_0)(h_0-1)+h_0^3(h_0^3+(h_0-1)^3)}\right]
\right]
\label{coupling2}
\end{eqnarray}
Let us note that the string couplings ($\xi_1(\phi)=\xi_1^{(0)} \e^{n\frac
{\phi}{\phi_0}}, \xi_2(\phi)=\xi_2^{(0)}
\e^{m\frac {\phi}{\phi_0}}$ with $m=n^2=4$) are generic to solution described
by (\ref{solution}); for other
couplings such a solution does not exist. We also note that
Eqs.(\ref{coupling1})
$\&$ (\ref{coupling2}) reduce
to the earlier obtained results in Ref.\cite{Nojiri:2005vv}
(see Refs.\cite{Calcagni:2005im, Neupane:2006dp,Carter:2005fu} on the related
theme)
   where similar investigations were carried out confining to
only second order curvature invariants in the action (\ref{act}).

There are several free parameters in the problem. In order to extract important
information from Eqs. (\ref{coupling1})
and (\ref{coupling2}), we proceed in the following manner. We fix $h_0$
corresponding to the observed value of
dark energy equation of state parameter $w_{DE}$ and impose the positivity
condition on the couplings $\xi_1^{(0)}$ and $\xi_2^{(0)}$
leading to allowed values of the parameter $\phi_0^2$.
In the absence of
coupling $\xi_2(\phi)$, it was shown in Ref.\cite{Nojiri:2005vv} that for a
given value of $h_0$ from the allowed interval, the parameter
$\phi_0$ takes a fixed value. Our model incorporates higher order curvature
corrections allowing a one
parameter  flexibility in the values of $\phi_0$. This gives rise to
comfortable choice of the equation of state
consistent with observations.

The three years WMAP data is analyzed in Ref.\cite{Spergel}. which
shows that the combined analysis of WMAP with supernova Legacy
survey (SNLS) constrains the dark energy equation of state $w_{DE}$
pushing it to wards the cosmological constant. The marginalized best
fit values of the equation of state parameter at 68$\%$ confidence
level are given by $-1.14\leq w_{DE} \leq -0.93$. In case of a prior
that universe is flat, the combined data gives  $-1.06 \leq w_{DE}
\leq -0.90 $. Our model can easily accommodate these values of
$w_{DE}$. For instance, in case of non-phantom (standard dark
energy) we find a one parameter family of  dark energy models with
$h_0 \simeq 40$ ($w_{DE}=-0.98$) corresponding to
     $\phi_0^2 >41$. Like wise, in case of phantom dark energy, we find that for
$h_0 \simeq -33$ ($w_{DE}=-1.02$),
the viable range of the parameter $\phi_0^2$ is given by $6<\phi_0^2<31.5$.
These values are consistent
with recent WMAP release and SNLS findings. 

We should mention that the observations
quoted above do not incorporate the dark energy perturbations which might severely
constrain the phantom dark energy cosmologies. The combined data (CMB+LSS+SNLS) then forces
the dark energy equation of state to vary as, $-1.001< w_{DE}< -0.875 $\cite{Spergel}. Our model
can easily incorporate these numerical values of $w_{DE}$ by constraining $h_0$ and $\phi_0^2$
similar to the case of non-clustering dark energy. A word of caution, the
evolution of dark energy perturbation across the phantom divide needs additional assumptions; 
a complete analysis should take into account the non-adiabatic perturbations 
which makes dark energy gravitationally stable\cite{CST}. 
\begin{table*}[t]
\begin{center}
\begin{tabular}{|c|c|c|c|c|c|}
\hline
Dark energy &  $h_0$ & $\phi_0^2$ & $w_{DE}$ & Observational constraint on
$w_{DE}$ & Constraint on $ w_{DE}$ with flatness prior \\
\hline
\hline
Non-phantom&40 &$\phi_0^2>41$ & $-0.98$ & &  \\
& & & &$-1.06^{+0.13}_{-0.08}$ & $-0.97^{+0.07}_{-0.09}$  \\

Phantom& $-33.33$ &$6<{\phi}_0^2<31.5 $ & $-1.02$ & &  \\
\hline
\end{tabular}
\end{center}
\caption[crit]{\label{crit} Observational constraints on (non-clustering) dark energy
equation of state $w_{DE}$
dictated by the combined analysis of WMAP+SNLS data\cite{Spergel} and the numerical values
of model parameters consistent with the observations.}
\end{table*}

\subsection{Stability of dark energy solution}

In what follows we shall examine the stability of the dark energy
solution (\ref{solution}) induced by purely stringy corrections. In
general the analytical treatment becomes intractable;
simplification, however, occurs in the limit of large $h_0$
corresponding to $w_{eff} \simeq -1$.

Let us consider the following situation of interest to us
\be 
\rho_m=0,~ \omega=\nu=1,~
  \xi_1(\phi)=\xi_1^{(0)}\e^{2\phi/\phi_0}\ ,\quad
\xi_2(\phi)=\xi_2^{(0)}\e^{4\phi/\phi_0}\ .
\label{P0}
\ee
In order to investigate the stability around the dark energy solution
defined by (\ref{solution}), we need a convenient set of variables
to cast the evolution equations into autonomous form. We now define
the variables which are suited to our problem.
\be 
\label{P1} 
{\cal X}\equiv \frac{\dot \phi}{H}\ ,\quad 
{\cal Y} \equiv \left(\dot H + H^2\right)^2 \xi_2(\phi)\ ,\quad 
{\cal Z} \equiv H^2 \xi_1(\phi)\ ,\quad \frac{d}{dN}\equiv \frac{1}{H}\frac{d}{dt}\ .
\ee
With this choice, the evolution equations acquire the autonomous form
\bea 
\label{P2} 
\frac{d{\cal X}}{dN}&=& - 2{\cal X} + \xi_1^{(0)}\left( - \frac{{\cal X}}{{\cal Z}} + \frac{48}{\phi_0} \right)
\left(\frac{{\cal Y}}{\xi_2^{(0)}}\right)^{1/2} 
+ \frac{96}{\phi_0}\frac{\xi_2^{(0)}}{\left(\xi_1^{(0)}\right)^2} {\cal Z}^2
+ \frac{96\xi_1^{(0)}{\cal Y}}{\phi_0 {\cal Z}}\left(\frac{{\cal Y}}{\xi_2^{(0)}}\right)^{1/2} \ ,\nn 
\frac{d{\cal Y}}{dN}&=& - \frac{1}{24\kappa^2} \frac{{\cal X}^2}{144} - \frac{2}{3\phi_0}{\cal Z}{\cal X} - 2{\cal Y} 
+ \frac{2\xi_1^{(0)}{\cal Y}}{3{\cal Z}}\left(\frac{{\cal Y}}{\xi_2^{(0)}}\right)^{1/2} 
+ \frac{5\xi_2^{(0)}}{3\left(\xi_1^{(0)}\right)^2} {\cal Z}^2\ ,\nn
\frac{d{\cal Z}}{dN} &=& \left( -2 + \frac{2}{\phi_0}{\cal X}\right){\cal Z} 
+ 2 \xi_1^{(0)}\left(\frac{{\cal Y}}{\xi_2^{(0)}}\right)^{1/2} \ . 
\eea
We have used the field equation (\ref{phieq}) and Eq.(\ref{Hddot}) 
for $\ddot{H}$ in deriving the above autonomous form of equations.
For our solution given by (\ref{solution}), we have
\be 
\label{P3} 
{\cal X}={\cal X}_0 \equiv \frac{\phi_0}{h_0}\ ,\quad 
{\cal Z}={\cal Z}_0 \equiv \frac{h_0^2 \xi_1^{(0)}}{t_1^2}\ ,\quad 
{\cal Y}={\cal Y}_0 = \frac{ \left( - h_0 + h_0^2 \right)^2 \xi_2^{(0)}}{t_1^4}\ . 
\ee
It can be checked that $({\cal X}_0,{\cal Y}_0,{\cal Z}_0)$ is a fixed point of (\ref{P2}).
We then consider small perturbations  around (\ref{P3}) or equivalently
around the original solution (\ref{solution})
\be 
\label{P4} 
{\cal X}={\cal X}_0 + \delta {\cal X}\ ,\quad 
{\cal Y}={\cal Y}_0 + \delta {\cal Y}\ ,\quad
{\cal Z}={\cal Z}_0 + \delta {\cal Z}\ .
\label{sp}
\ee
Substituting (\ref{sp}) in (\ref{P2}) and retaining the linear terms
in perturbations, we find
\bea 
\label{P5} 
\frac{d}{dN}\left(\begin{array}{c} \delta {\cal X} \\ \delta {\cal Y} \\ \delta {\cal Z}
\end{array}\right)
= M \left(\begin{array}{c} \delta {\cal X} \\ \delta {\cal Y} \\ \delta {\cal Z}
\end{array}\right) \ .
\eea
Here $M$ is $3\times 3$-matrix perturbation matrix whose components
are given by
\bea 
\label{P6} 
M_{11} &=&  -2 + \frac{- 1 + h_0}{h_0} \ ,\nn 
M_{12} &=& - \frac{\phi_0 t_1^2}{2h_0^3} +
\frac{24\xi_1^{(0)}t_1^2}{\phi_0\xi_2^{(0)}\left( - h_0 + h_0^2 \right)} 
+ \frac{144\left(-1 + h_0\right)}{\phi_0 h_0 } \ ,\nn
M_{13} &=& \frac{\phi_0 t_1^2 \left( -1 + h_0 \right) }{h_0^4 \xi_1^{(0)}} 
+ \frac{192\xi_2^{(0)}h_0^2}{\xi_1^{(0)}\phi_0 t_1^2}
 - \frac{96 \left( -1 + h_0 \right)^3 \xi_2^{(0)}}{\phi_0 h_0 \xi_1^{(0)} t_1^2} \ ,\nn
M_{21} &=& \frac{1}{72} - \frac{2h_0^2 \xi_1^{(0)}}{3\phi_0 t_1^2} \ ,\nn 
M_{22} &=& - 2 - \frac{\left( - 1 + h_0 \right)}{h_0} \ ,\nn
M_{23} &=& - \frac{2}{3h_0} - \frac{2 \left( - 1 + h_0 \right)^3 \xi_2^{(0)}}{3h_0 \xi_1^{(0)} t_1^2} 
+ \frac{10 h_0^2 \xi_2^{(0)}}{3 \xi_1^{(0)}t_1^2} \ ,\nn 
M_{31} &=& \frac{2h_0^2 \xi_1^{(0)}}{\phi_0 t_1^2} \ ,\nn 
M_{32} &=& \frac{\xi_1^{(0)} t_1^2}{\xi_2^{(0)}\left(- h_0 + h_0^2 \right)} \ ,\nn 
M_{33} &=& -2 + \frac{2}{h_0}\ .
\eea
Stability of the fixed point(s) depends upon the nature eigenvalues
of the perturbation matrix $M$. If there is an eigenvalue whose real
part is positive, the system becomes unstable. Here for simplicity,
we only consider the case of $h_0\to \pm \infty$, which corresponds
to the limit of $w_{\rm eff}\sim -1$. In the case, we find
\be 
\label{P6b} 
\frac{\xi_1^{(0)}}{t_1^2}\to \frac{1}{40 h_0^5}\ ,\quad 
\frac{\xi_2^{(0)}}{t_1^4}\to - \frac{1}{32 h_0}\ , 
\ee 
and the eigenvalue equation is given by 
\be 
\label{P7} 
0=F(\lambda)
\equiv -\lambda^3 - 6\lambda^2 - \frac{h_0^3}{40\phi_0} \lambda - \frac{7 h_0^3}{40\phi_0} \ . 
\ee 
The values of $\lambda$ satisfying $F(\lambda)=0$ give eigenvalues of $M$. 
The solutions of (\ref{P7}) is given by 
\be 
\label{P8} 
\lambda=\lambda_\pm \equiv 
\pm \frac{\left| h_0 \right| \sqrt{- h_0}}{\phi_0 \sqrt{40}} 
+ {\cal O}\left(\left|h_0\right|\right)\ , \quad 
\lambda = \lambda_0 \equiv -7 + {\cal O}\left(\left|h_0\right|^{-1/2}\right)\ . 
\ee 
When $h_0<0$, the mode corresponding to $\lambda_+$
$\left(\lambda_-\right)$ becomes stable (unstable). Since
$\lambda_\pm$ are pure imaginary when $h_0>0$, the corresponding
modes become stable in this case. On the other hand, the mode corresponding to
$\lambda_0$ is always stable. Thus, the non-phantom dark energy
solution (\ref{solution}) induced by string corrections to
Einstein gravity is stable. Such a solution exists in presence
of a dynamically evolving field $\phi$ with $V(\phi)=0$ coupled to Riemann invariants
with couplings dictated by string theory. Dark energy can be 
realized in a variety of scalar field models by appropriately 
choosing the field potential. It is really interesting that we can obtain dark energy solution
in string model without recourse to a scalar field potential.
 
Let us compare the results with
those obtained in \cite{Nojiri:2005vv}, where $\xi_2=0$ but $V(\phi)\neq 0$. The dark energy
solution studied 
in Ref.\cite{Nojiri:2005vv}, was shown to be stable when $h_0>0$ but unstable
for $h_0>0$. The present investigations include $\xi_2$ and $V=0$ which makes our model different 
from Ref.\cite{Nojiri:2005vv}; it is therefore not surprising that
our results differ from Ref.\cite{Nojiri:2005vv}. Since $h_0>0$
corresponds to the quintessence phase and $h<0$ to the phantom, the
solution in the model (with $\xi_0$ and $V=0$) is
stable in the quintessence phase but unstable in the phantom phase.
We should notice that the approximation we used to check the stability
works fine for any generic
value of $h_0$. For instance, $5 < h_0 < -667$ which corresponds to
the variation of $w_{DE}$ in case the dark energy perturbations are taken into account.
We also carried out numerical verification of our results.  


\section{The late-time cosmology in scalar-Gauss-Bonnet gravity}

A number of scalar field models have recently been investigated
in connection with dark energy (see Ref.\cite{CST} for details). The cosmological viability
of these constructs depends upon how well the Hubble parameter
predicted by them compares with observations. One could also follow the reverse
route and construct the Lagrangian using the observational input; such a scheme
might help in the search of best fit models of dark energy\cite{CST}. In what follows we shall 
describe how the reconstruction program is implemented in presence of higher order
string curvature corrections. 

\subsection{The reconstruction of scalar-Gauss-Bonnet gravity}

In this section it will be shown how scalar-Gauss-Bonnet gravity may be
reconstructed for any requested cosmology using the method \cite{e}
developed in the scalar-tensor theory.  We limit here only by Gauss-Bonnet term
order (by technical reasons)  but there is no principal problem to include
higher order terms studied in previous section.
It is interesting that the principal possibility appears to reconstruct
the scalar-Gauss-Bonnet gravity for any (quintessence, cosmological constant or
phantom) dark energy universe. The last possibility seems to be quite
attractive due to the fact \cite{Nojiri:2005vv},
    that the phantom universe could be realized in
the scalar-Gauss-Bonnet gravity without introducing ghost scalar field.
In this section, we show that in  scalar-Gauss-Bonnet gravity, any cosmology,
including
phantom cosmology, could be realized by properly choosing the potential and the
coupling to the Gauss-Bonnet invariant with the canonical scalar.

The starting action is
\be
\label{GBany1}
S=\int d^4 x \sqrt{-g}\left[ \frac{R}{2\kappa^2} - \frac{1}{2}\partial_\mu \phi
\partial^\mu \phi - V(\phi) - \xi_1(\phi) G \right]\ .
\ee
Here $G$ is the Gauss-Bonnet invariant $G\equiv {\cal L}_c^{(1)}$ (\ref{Lc1})
and the scalar field $\phi$ is canonical in (\ref{GBany1}).
As in previous section, it is natural to assume the FRW universe (\ref{FRW})
with $N(t)=1$ and the scalar field $\phi$ only depending on $t$.
The FRW equations look like:
\bea
\label{GBany4}
0&=& - \frac{3}{\kappa^2}H^2 + \frac{1}{2}{\dot\phi}^2 + V(\phi) + 24 H^3
\frac{d \xi_1(\phi(t))}{dt}\ ,\\
\label{GBany5}
0&=& \frac{1}{\kappa^2}\left(2\dot H + 3 H^2 \right) + \frac{1}{2}{\dot\phi}^2
- V(\phi)
    - 8H^2 \frac{d^2 \xi_1(\phi(t))}{dt^2} - 16H \dot H
\frac{d\xi_1(\phi(t))}{dt}
    - 16 H^3 \frac{d \xi_1(\phi(t))}{dt}\ .
\eea
and scalar field equation:
\be
\label{GBany6}
0=\ddot \phi + 3H\dot \phi + V'(\phi) + \xi_1'(\phi) G\ .
\ee

Combining (\ref{GBany4}) and (\ref{GBany5}), one gets
\be
\label{GBany7}
0=\frac{2}{\kappa^2}\dot H + {\dot\phi}^2 - 8H^2 \frac{d^2
\xi_1(\phi(t))}{dt^2}
    - 16 H\dot H \frac{d\xi_1(\phi(t))}{dt} + 8H^3 \frac{d\xi_1(\phi(t))}{dt}
=\frac{2}{\kappa^2}\dot H + {\dot\phi}^2 -
8a\frac{d}{dt}\left(\frac{H^2}{a}\frac{d\xi_1(\phi(t))}{dt}\right)\ .
\ee
Eq.(\ref{GBany7}) can be solved with respect to $\xi_1(\phi(t))$ as
\be
\label{GBany8}
\xi_1(\phi(t))=\frac{1}{8}\int^t dt_1 \frac{a(t_1)}{H(t_1)^2} \int^{t_1}
\frac{dt_2}{a(t_2)}
\left(\frac{2}{\kappa^2}\dot H (t_2) + {\dot\phi(t_2)}^2 \right)\ .
\ee
Combining (\ref{GBany4}) and (\ref{GBany8}), the scalar potential $V(\phi(t))$
is:
\be
\label{GBany9}
V(\phi(t)) = \frac{3}{\kappa^2}H(t)^2 - \frac{1}{2}{\dot\phi (t)}^2  - 3a(t)
H(t) \int^t \frac{dt_1}{a(t_1)}
\left(\frac{2}{\kappa^2}\dot H (t_1) + {\dot\phi(t_1)}^2 \right)\ .
\ee
We now identify $t$ with $f(\phi)$ and $H$ with $g'(t)$ where $f$ and $g$
are some unknown functions in analogy with Ref.\cite{e} since we know this
leads to the solution of the FRW equations subject to existence of
such functions.
Then we consider the model where $V(\phi)$ and $\xi_1(\phi)$ may be expressed
in terms of two functions $f$ and $g$ as
\bea
\label{GBany10b}
V(\phi) &=& \frac{3}{\kappa^2}g'\left(f(\phi)\right)^2 - \frac{1}{2f'(\phi)^2}
    - 3g'\left(f(\phi)\right) \e^{g\left(f(\phi)\right)} \int^\phi d\phi_1
f'(\phi_1 ) \e^{-g\left(f(\phi_1)\right)} \times
\left(\frac{2}{\kappa^2}g''\left(f(\phi_1)\right) + \frac{1}{f'(\phi_1 )^2}
\right)\ ,\nn
\xi_1(\phi) &=& \frac{1}{8}\int^\phi d\phi_1 \frac{f'(\phi_1)
\e^{g\left(f(\phi_1)\right)} }{g'(\phi_1)^2}
\int^{\phi_1} d\phi_2  f'(\phi_2) \e^{-g\left(f(\phi_2)\right)}
\left(\frac{2}{\kappa^2}g''\left(f(\phi_2)\right) + \frac{1}{f'(\phi_2)^2}
\right)\ .
\eea
By choosing $V(\phi)$ and $\xi_1(\phi)$ as (\ref{GBany10b}), we can easily find
the following solution for Eqs.(\ref{GBany4}) and (\ref{GBany5}):
\be
\label{GBany11b}
\phi=f^{-1}(t)\quad \left(t=f(\phi)\right)\ ,\quad
a=a_0\e^{g(t)}\ \left(H= g'(t)\right)\ .
\ee
We can straightforwardly check the solution (\ref{GBany11b}) satisfies the
field equation (\ref{GBany6}).

Hence, any cosmology expressed as $H=g(\phi)$ in the model (\ref{GBany1}) with
(\ref{GBany10b}) can be realized, including
    the model exhibiting the transition from non-phantom phase to phantom phase
without introducing
the scalar field with wrong sign  kinetic term.


In the Einstein gravity, the FRW equations are given by
\be
\label{GBany12}
0= - \frac{3}{\kappa^2}H^2 + \rho\ ,\quad
0= \frac{1}{\kappa^2}\left(2\dot H + 3 H^2 \right) + p\ .
\ee
Here $\rho$ and $p$ are total energy density and pressure in the universe.
By comparing (\ref{GBany12}) with (\ref{GBany11b}) we find the effective energy
density $\tilde\rho$
and the pressure $\tilde p$ are given
\be
\label{GBany13b}
\tilde\rho= \frac{3}{\kappa^2}g'(t)^2\ ,\quad
\tilde p= -\frac{3}{\kappa^2}g'(t)^2 - \frac{2}{\kappa^2}g''(t)\ .
\ee
Since $t={g'}^{-1}\left(\left(\kappa\right)\sqrt{\rho/3}\right)$, we obtain
the following effective equation of the state (EoS):
\be
\label{GBany14}
\tilde p=-\tilde \rho -
\frac{2}{\kappa^2}g''\left({g'}^{-1}\left(\kappa\sqrt{\frac{\rho}{3}}\right)\right)\
,
\ee
which contains all the cases where the EoS is given by $p=w(\rho)\rho$.
Furthermore, since ${g'}^{-1}$ could NOT be always a single-valued function,
Eq.(\ref{GBany14}) contains more general EoS given by
\be
\label{GBany15}
0=F\left(\tilde\rho,\tilde p\right)\ .
\ee
This shows the equivalence between scalar-tensor and ideal fluid
descriptions.

Let us come back now to scalar-Gauss-Bonnet gravity.
It is not difficult to extend the above formulation to include matter
with constant EoS parameter $w_m\equiv p_m / \rho_m$.
Here $\rho_m$ and $p_m$ are energy density and pressure of the matter.
Then, instead of (\ref{GBany4}) and (\ref{GBany5}) the FRW equations are
\bea
\label{GBany16}
0&=& - \frac{3}{\kappa^2}H^2 + \frac{1}{2}{\dot\phi}^2 + V(\phi) + \rho_m
+ 24 H^3 \frac{d \xi_1(\phi(t))}{dt}\ ,\\
\label{GBany17}
0&=& \frac{1}{\kappa^2}\left(2\dot H + 3 H^2 \right) + \frac{1}{2}{\dot\phi}^2
- V(\phi) + p_m
    - 8H^2 \frac{d^2 \xi_1(\phi(t))}{dt^2} - 16H \dot H
\frac{d\xi_1(\phi(t))}{dt}
    - 16 H^3 \frac{d \xi_1(\phi(t))}{dt}\ .
\eea
The energy conservation law
\be
\label{GBany18}
\dot\rho_m + 3H\left(\rho_m + p_m\right)=0\ ,
\ee
gives \be
\label{GBany19}
\rho_m=\rho_{m0} a^{-3(1+w_m)}\ ,
\ee
with a constant $\rho_{m0}$.
Instead of (\ref{GBany10b}), if we consider the model with
\bea
\label{GBany20b}
V(\phi) &=& \frac{3}{\kappa^2}g'\left(f(\phi)\right)^2 - \frac{1}{2f'(\phi)^2}
    - 3g'\left(f(\phi)\right) \e^{g\left(f(\phi)\right)} \int^\phi d\phi_1
f'(\phi_1)\e^{-g\left(f(\phi_1)\right)}
\left(\frac{2}{\kappa^2}g''\left(f(\phi_1)\right) + \frac{1}{2f'(\phi_1 )^2}
\right. \nn
&& \left. + (1+w_m)g_0\e^{-3(1+w_m)g\left(f(\phi_1)\right)}\right) \ ,\nn
\xi_1(\phi) &=& \frac{1}{8}\int^\phi d\phi_1 \frac{f'(\phi_1)
\e^{g\left(f(\phi_1)\right)} }{g'(\phi_1)^2}
\int^{\phi_1} d\phi_2  f'(\phi_2) \e^{-g\left(f(\phi_2)\right)}
\left(\frac{2}{\kappa^2}g''\left(f(\phi_2)\right) \right. \nn
&& \left. + \frac{1}{2f'(\phi_2)^2} +
(1+w_m)g_0\e^{-3(1+w_m)g\left(f(\phi_2)\right)} \right)\ ,
\eea
we re-obtain the solution  (\ref{GBany11b}) even if  the matter is
included.
However,
a constant $a_0$ is given by
\be
\label{GBany21}
a_0=\frac{g_0}{\rho_0}\ .
\ee

One can consider some explicit examples \cite{e}:
\be
\label{GBany22}
t=f(\phi)=\frac{\phi}{\phi_0}\ ,\quad g(t)=h_0\ln \frac{t}{t_s - t}\ ,
\ee
which gives
\be
\label{GBany23}
H=h_0\left(\frac{1}{t} + \frac{1}{t_s - t}\right)\ ,\quad
\dot H = \frac{h_0 t_s \left(2t - t_s\right)}{t^2\left(t_s - t\right)^2}\ .
\ee
Then the universe is in non-phantom phase when $t<t_s/2$ and in phantom phase
when $t>t_s/2$.
There is also a Big Rip singularity at $t=t_s$.
Especially in case $w_m=0$ (that is, matter is dust) and $h_0=2$, we
reconstruct the scalar-Gauss-Bonnet gravity with following potentials:
\bea
\label{GBany24}
V(\phi)&=& \frac{6\phi_0 \phi_s}{\kappa^2\phi\left(\phi_s - \phi \right)} -
\frac{1}{2}\phi_0^2
    - \frac{4\phi_0^2\phi_s \phi}{\left(\phi_s - \phi\right)^3}\left[
\frac{4}{\kappa^2}\left(\frac{\phi_s^2}{3\phi^3} - \frac{\phi_s}{\phi^2}\right)
    - \frac{\phi_s^2}{\phi} - 2\phi_s \ln \frac{\phi}{\phi_s} + \phi \right. \nn
&& \left. + \frac{g_0}{\phi_0^2}\left(- \frac{\phi_s^8}{7\phi^7} +
\frac{4\phi_s^7}{3\phi^6}
    - \frac{28\phi_s^6}{5\phi^5}
+ \frac{14\phi_s^5}{\phi^4} - \frac{70\phi_s^4}{3\phi^3} +
\frac{28\phi_s^3}{\phi^2}
    - \frac{28\phi_s^2}{\phi} - 8\phi_s \ln \frac{\phi}{\phi_s} + \phi\right) +
c_1 \right]\ ,\nn
\xi_1(\phi)&=& \frac{1}{32\phi_0^2\phi_s^2}\left[
\frac{4}{\kappa^2}\left(\frac{\phi_s^2\phi^2}{6}
    - \frac{\phi_s\phi^3}{3}\right)
    - \frac{\phi_s^2 \phi^4}{4} - 2\phi_s \phi^5\left(\frac{1}{5}\ln
\frac{\phi}{\phi_s} - \frac{1}{25}\right)
+ \frac{\phi^6}{6}
+\frac{g_0}{\phi_0^2}\left(\frac{\phi_s^8}{14\phi^2} - \frac{4\phi_s^7}{3\phi}
    - \frac{28\phi_s^6}{5}\ln \frac{\phi}{\phi_s} \right.\right. \nn
&& \left.\left. + 14\phi_s^5 \phi - \frac{35\phi_s^4 \phi^2}{3} +
\frac{28\phi_s^3\phi^3}{3}
    - 7\phi_s^2\phi^4 - 8 \phi_s \phi^5\left(\frac{1}{5}\ln\frac{\phi}{\phi_s} -
\frac{1}{25}\right)
+ \frac{\phi^6}{6}\right) + \frac{c_1 \phi^5}{5} + c_2 \right]\ .
\eea
Here $\phi_s\equiv \phi_0 t_s$ and $c_1$, $c_2$ are constants of the
integration.

Another example, without matter ($g_0=0$), is \cite{osc}
\be
\label{GBany25}
g(t) = h_0\left( t + \frac{\cos \theta_0}{\omega}\sin \omega t \right)\ ,\quad
f^{-1}(t)= \phi_0 \sin\frac{\omega t}{2}\ .
\ee
Here $h_0$, $\theta_0$, $\omega$, and $\phi_0$ are constants.
This leads to reconstruction of scalar-Gauss-Bonnet gravity with \bea
\label{GBany26}
V(\phi)&=&\frac{3h_0}{\kappa^2}\left(1 + \cos\theta_0 -
\frac{2\cos\theta_0}{\phi_0^2}\phi^2\right)
    - \frac{\phi_0^2 \omega^2}{8}\left( 1 -
\frac{\phi^2}{\phi_0^2}\right)^{1/2}\
,\nn
\xi_1(\phi)&=& - \frac{\omega\phi_0}{32h_0^3}\int^\phi d\phi_1 \left( 1 -
\frac{\phi_1^2}{\phi_0^2}\right)^{-1/2}
\left( 1 + \cos\theta_0 - \frac{2\cos\theta_0}{\phi_0^2}\phi_1^2 \right)^{-2}\
,
\eea
Then from Eq.(\ref{GBany25}) we find
\be
\label{GBany27}
H=h_0\left( 1 + \cos \theta_0 \cos \omega t \right)\geq 0 \ ,\quad
\dot H = - h_0 \omega \cos \theta_0 \sin \omega t\ ,
\ee
Then the Hubble rate $H$ is oscillating but since $H$ is positive,
the universe continues to expand and if $h_0 \omega \cos \theta_0>0$, the
universe is in non-phantom (phantom)
phase when $2n \pi < \omega t < \left(2n + 1\right) \pi$ $\left(\left(2n -
1\right) \pi < \omega t < 2n \pi\right)$
with integer $n$. Thus, the oscillating late-time cosmology in string-inspired
gravity may be easily constructed.

One more example is \cite{osc}
\be
\label{GBany28}
g(t)=H_0 t - \frac{H_1}{H_0}\ln \cosh H_0 t\ .
\ee
Here we assume $H_0>H_1>0$. Since
\be
\label{GBany29}
H=g'(t)=H_0 - H_1\tanh H_0 t\ ,\quad
\dot H=g''(t)=- \frac{H_0H_1}{\cosh^2 H_0 t}<0\ ,
\ee
when $t\to \pm \infty$, the universe becomes asymptotically deSitter space,
where $H$ becomes a constant
$H\to H_0 \mp H_1$ and therefore the universe is accelerating.
When $t=0$, we find
\be
\label{GBany30}
\frac{\ddot a}{a}=\dot H + H^2 = -H_1 H_0 + H_0^2 < 0\ ,
\ee
therefore the universe is decelerating.
Then the universe is accelerating at first, turns to be decelerating, and
after that universe becomes
accelerating again. As $\dot H$ is always negative, the universe is in
non-phantom phase.
Furthermore with the choice
\be
\label{GBany31}
w_m=0\ ,\quad t=f(\phi)= \frac{1}{H_0}\tan \left(\frac{\kappa H_0}{2\sqrt{2}
H_1}\phi\right) \ ,
\ee
we find the corresponding scalar-Gauss-Bonnet gravity
\bea
\label{GBany32}
V(\phi)&=&\frac{3}{\kappa^2}\left(H_0 - H_1 \tanh \varphi \right)^2 -
\frac{H_1}{\sqrt{2} \kappa^2 \cosh^2 \varphi} \nn
&& - \frac{12g_0}{H_0}\left(H_0 - H_1 \tanh \varphi
\right)\left(1+\e^{2\varphi}\right)
\left[ 2\varphi - \ln \left(1+\e^{2\varphi}\right) +
\frac{5}{6\left(1+\e^{2\varphi}\right)}
+ \frac{5}{6\left(1+\e^{2\varphi}\right)^2} +
\frac{2}{6\left(1+\e^{2\varphi}\right)^3}\right]\ ,\nn
\xi_1(\phi)&=& \frac{g_0}{2H_0}\int^\varphi d\varphi' \frac{1+\e^{2\varphi'}}{
\left(H_0 - H_1 \tanh \varphi' \right)^2}
\left[ 2\varphi' - \ln \left(1+\e^{2\varphi'}\right) +
\frac{5}{6\left(1+\e^{2\varphi'}\right)}
+ \frac{5}{6\left(1+\e^{2\varphi'}\right)^2} +
\frac{2}{6\left(1+\e^{2\varphi'}\right)^3}\right] .
\eea
Here
\be
\label{GBany33}
\varphi\equiv \tan \left(\frac{\kappa H_0}{2\sqrt{2} H_1}\phi\right) \ .
\ee

Although it is difficult to give the explicit forms of $V(\phi)$ and
$\xi_1(\phi)$, we may also consider the following example \cite{e}:
\be
\label{GBA1}
g(t)=h_0 \left(\frac{t^4}{12} - \frac{t_1+t_2}{6}t^3 + \frac{t_1 +
t_2}{2}t^2\right)\ ,\quad
\left(3t_1>t_2>t_1>0\ , \quad h_0>0\right)\ .
\ee
Here $h_0$, $t_1$, $t_2$ are constants.
Hence, Hubble rate is
\be
\label{GBA2}
H(t) = h_0 \left(\frac{t^3}{3} - \frac{t_1 + t_2}{2}t^2 + t_1 t_2 t \right)\
,\quad
\dot H (t) = h_0 \left(t-t_1\right)\left(t - t_2\right)\ .
\ee
Since $H>0$ when $t>0$ and $H<0$ when $t<0$, the radius of the universe
$a=a_0\e^{g(t)}$ has a minimum when $t=0$.
 From the expression of $\dot H$ in (\ref{GBA2}), the universe is in phantom
 phase $\left(\dot H>0\right)$
when $t<t_1$ or $t>t_2$, and in non-phantom phase $\left(\dot H<0\right)$ when
$t_1<t<t_2$ (for other string-inspired models with similar cosmology, see
for instance \cite{ar}).
Then we may identify the period $0<t<t_1$ could correspond to the inflation and
the period $t>t_2$ to the present acceleration of the universe (this is
similar in spirit to unification of the inflation with the acceleration
suggested in other class of modified gravities in ref.\cite{prd}).
If we define effective EoS parameter $w_{\rm eff}$ as
\be
\label{FRW3k}
w_{\rm eff}=\frac{p}{\rho}= -1 - \frac{2\dot H}{3H^2}\ ,
\ee
we find $w_{\rm eff}\to -1$ in the limit $t\to + \infty$.
Although it is difficult to find the explicit forms of $V(\phi)$ and
$\xi_1(\phi)$, one might give the rough forms by
using the numerical calculations. From the expression of $V(\phi)$ and
$\xi_1(\phi)$ in (\ref{GBany10b}),
if $f(\phi)$ is properly given, say as $t=f(\phi)=\phi/\phi_0$ with constant
$\phi_0$, there cannot happen any
singularity in $V(\phi)$ and $\xi_1(\phi)$ even if $t=t_1$ or $t=t_1$, which
corresponds to the transition
between phantom and non-phantom phases. Then the model (\ref{GBany1}) could
exhibit the smooth transition
between phantom and non-phantom phases.

The next example is
\be
\label{GBany37}
g(t)=h_0\ln \frac{t}{t_0}\ ,\quad
t=f(\phi)=t_0\e^{\frac{\phi}{\phi_0}}\ .
\ee
Since
\be
\label{GBany38}
H=\frac{h_0}{t}\ ,
\ee
we have a constant effective EoS parameter:
\be
\label{GBany39}
w_{\rm eff}=-1 + \frac{2}{3h_0}\ .
\ee
Eqs.(\ref{GBany37}) give
\bea
\label{GBany40}
V(\phi)&=& -\frac{1}{\left(h_0 +
1\right)t_0^2}\left(\frac{3h_0^2\left(1-h_0\right)}{\kappa^2}
+ \frac{\phi_0^2}{2}\left(1-5h_0\right)\right) \e^{-\frac{2\phi}{\phi_0}}
+ \frac{3h_0\left(1+w_m\right)g_0}{\left( 4+3w_m\right)h_0 - 1}
\e^{-\frac{3\left(1+w_m\right)h_0\phi}{\phi_0}}\ , \nn
\xi_1(\phi)&=& -\frac{t_0^2}{16h_0^2\left(h_0 + 1\right)}\left( -
\frac{2h_0}{\kappa^2} + \phi_0^2
\right)\e^{\frac{2\phi}{\phi_0}} \nn
&& + \frac{1}{8}\left\{3\left(1+w_m\right)h_0 - 4\right\}^{-1}
\left\{\left(4+3w_m\right)h_0 - 1\right\}^{-1}
\left(1+w_m\right)g_0 t_0^4 \e^{-\frac{\left\{3\left(1+w_m\right)h_0 -
4\right\}\phi}{\phi_0}}\ .
\eea
Thus, there appear exponential functions, which are typical in string-inspired
gravity.

As clear from (\ref{GBany38}), if $h_0>1$, the universe is in quintessence
phase, which corresponds to
$-1/3<w_{\rm eff}<-1$ in (\ref{GBany37}). If $h_0<0$, the universe is in
phantom phase with $w_{\rm eff}<-1$.
In phantom phase, we choose $t_0$ to be negative and our universe corresponds
to negative $t$, or if we
shift the time coordinate $t$ as $t\to t - t_s$, with a constant $t_s$, $t$
should be less than $t_s$.

The model \cite{Nojiri:2005vv} corresponds to $g_0=0$ in (\ref{GBany40}).
    In the notations of ref.\cite{Nojiri:2005vv}, $t_0=t_1$,
$V(\phi)=V_0\e^{-\frac{2\phi}{\phi_0}}$, and
$f(\phi)=f_0\e^{\frac{2\phi}{\phi_0}}=-\xi_1(\phi)$. Then from the expression
    (\ref{GBany40}), one gets
\be
\label{GBany40a}
V_0= -\frac{1}{\left(h_0 +
1\right)t_0^2}\left(\frac{3h_0^2\left(1-h_0\right)}{\kappa^2}
+ \frac{\phi_0^2}{2}\left(1-5h_0\right)\right)\ , \quad
f_0= \frac{t_0^2}{16h_0^2\left(h_0 + 1\right)}\left( - \frac{2h_0}{\kappa^2} +
\phi_0^2 \right) \ ,
\ee
which is identical (after replacing $t_0$ with $t_1$) with (16) in
\cite{Nojiri:2005vv}.

Thus, we demonstrated that arbitrary late-time cosmology (from specific
quintessence or phantom to oscillating cosmology) may be produced by
scalar-Gauss-Bonnet gravity with scalar potentials defined by such cosmology.
The reconstruction of string-inspired gravity may be always done. Moreover, one
can extend this formulation to include the higher order terms in low-energy
string effective action.

\subsection{The relation with modified Gauss-Bonnet gravity}

In this section we show that scalar-Gauss-Bonnet gravity may be transformed
to another form of modified Gauss-Bonnet gravity where no scalars present.
In addition, the formulation may be extended to include higher order terms
too.
Starting from (\ref{GBany1}), one may redefine the scalar field $\phi$ by
$\phi=\epsilon
\varphi$.
The action takes the following form
\be
\label{GBany41}
S=\int d^4 x \sqrt{-g}\left[ \frac{R}{2\kappa^2} -
\frac{\epsilon^2}{2}\partial_\mu \phi \partial^\mu \phi
    - \tilde V(\varphi) - \tilde\xi_1(\varphi) G \right]\ .
\ee
Here
\be
\label{GBany42}
\tilde V(\varphi) \equiv V(\epsilon\varphi)\ ,\quad
\tilde\xi_1(\varphi) \equiv \xi_1(\epsilon\varphi)\ .
\ee
If a proper limit of $\epsilon\to 0$ exists, the action (\ref{GBany41}) reduces
to
\be
\label{GBany43}
S=\int d^4 x \sqrt{-g}\left[ \frac{R}{2\kappa^2} - \tilde V(\varphi) -
\tilde\xi_1(\varphi) G \right]\ .
\ee
Then  $\varphi$ is an auxiliary field. By the variation of
$\varphi$, we find
\be
\label{GBany44}
0={\tilde V}'(\varphi) - {\tilde\xi_1}'(\varphi)G\ ,
\ee
which may be solved with respect to $\varphi$ as
\be
\label{GBany45}
\varphi=\Phi(G)\ .
\ee
Substituting (\ref{GBany46}) into the action (\ref{GBany43}), the
$F(G)$-gravity follows \cite{GB}:
\be
\label{GBany46}
S=\int d^4 x \sqrt{-g}\left[ \frac{R}{2\kappa^2} - F(G)\right]\ ,\quad
F(G)\equiv \tilde V\left(\Phi(G)\right) - \tilde\xi_1\left(\Phi(G)\right) G\ .
\ee
For example, in case of (\ref{GBany37}), in $\epsilon\to 0$ limit
after redefining
$\phi=\epsilon\varphi$ and $\phi_0=\epsilon\varphi_0$, $V(\phi)$ and
$\xi_1(\phi)$ reduce to
\bea
\label{GBany47}
\tilde V(\varphi)&=& \frac{3h_0^2\left(h_0 - 1\right)}{\left(h_0 +
1\right)t_0^2\kappa^2}\e^{-\frac{2\varphi}{\varphi_0}}
+ \frac{3h_0\left(1+w_m\right)g_0}{\left( 4+3w_m\right)h_0 - 1}
\e^{-\frac{3\left(1+w_m\right)h_0\phi}{\phi_0}}\ , \nn
\tilde\xi_1(\varphi)&=& \frac{t_0^2}{8h_0\left(h_0 +
1\right)\kappa^2}\e^{\frac{2\phi}{\phi_0}}
+ \frac{1}{8}\left\{3\left(1+w_m\right)h_0 - 4\right\}^{-1}
\left\{\left(4+3w_m\right)h_0 - 1\right\}^{-1}
\left(1+w_m\right)g_0 t_0^4 \e^{-\frac{\left\{3\left(1+w_m\right)h_0 -
4\right\}\phi}{\phi_0}}\ .
\eea
The solution corresponding to (\ref{GBany37}) is:
\be
\label{GBany48}
g(t)=h_0\ln \frac{t}{t_0}\ ,\quad \varphi=\varphi_0\ln \frac{t}{t_0} .
\ee
If we further consider the case $g_0=0$, Eq.(\ref{GBany44}) gives
\be
\label{GBany49}
\e^{-\frac{4\varphi}{\varphi_0}}=\frac{t_0^4}{24 h_0^3 \left(h_0-1\right)}G\ .
\ee
Eq.(\ref{GBany49}) could have a meaning only when $h_0>1$ or $h_0<0$ if
$G$ is positive.
In this situation
\be
\label{GBany50}
F(G)=A_0 G^{1/2}\ ,\quad
A_0\equiv \frac{1}{2\left(1+h_0\right)\kappa^2}\sqrt{\frac{3\left(h_0 -
1\right)h_0}{2}}\ .
\ee
The above model has been discussed in \cite{GB}.
Actually, in
\cite{GB}
the following type of the action has been considered:
\be
\label{GB1}
S=\int d^4 x\sqrt{-g}\left(\frac{1}{2\kappa^2}R + F(G)\right)\ .
\ee
In case that $F(G)$ is given by (\ref{GBany50}), in terms of \cite{GB},
$A_0=f_0$.
Hence, $A_0$  (\ref{GBany50}) coincides with the
    Eq.(26) of \cite{GB}.

As a further generalization, we may also consider the
string-inspired theory of second section where next order term  is
coupled with scalar field:
\be
\label{Eb1}
S=\int d^4 x \sqrt{-g}\Bigl[ \frac{R}{2\kappa^2} - \frac{1}{2}\partial_\mu \phi
\partial^\mu \phi
    - V(\phi)  - \xi_1(\phi) G + \xi_2(\phi) {\cal L}_c^{(2)}\Bigr]\ .
\ee
As in (\ref{GBany41}), we may redefine the scalar field $\phi$ by
$\phi=\epsilon \varphi$.
If a proper limit of $\epsilon\to 0$ exists, the action (\ref{Eb1}) reduces to
\be
\label{Eb2}
S=\int d^4 x \sqrt{-g}\left[ \frac{R}{2\kappa^2} - \tilde V(\varphi)
    - \tilde\xi_1(\varphi) G + \tilde\xi_2(\varphi){\cal L}_c^{(2)} \right]\ .
\ee
Here
\be
\label{Eb3}
\tilde\xi_2 = \lim_{\epsilon\to 0}\xi_2(\epsilon\varphi)\ .
\ee
Then $\varphi$ could be regarded as an auxiliary field and one gets
\be
\label{Eb4}
0={\tilde V}'(\varphi) - {\tilde\xi_1}'(\varphi)G + {\tilde\xi_2}'{\cal
L}_c^{(2)} \ ,
\ee
which may be solved with respect to $\varphi$ as
\be
\label{Eb5}
\varphi=\Psi\left(G,{\cal L}_c^{(2)}\right)\ .
\ee
Substituting (\ref{Eb5}) into the action (\ref{Eb2}), we obtain
$F(G,{\cal L}_c^{(2)})$-gravity theory:
\be
\label{Eb6}
S=\int d^4 x \sqrt{-g}\left[ \frac{R}{2\kappa^2} - F(G,{\cal
L}_c^{(2)})\right]\ ,\quad
F(G)\equiv \tilde V\left(\Phi(G,{\cal L}_c^{(2)})\right) -
\tilde\xi_1\left(\Phi(G,{\cal L}_c^{(2)})\right)
G
+ \tilde\xi_2\left(\Phi(G,{\cal L}_c^{(2)})\right){\cal L}_c^{(2)}\ .
\ee
In case of the string-inspired gravity:
\be
\label{E7}
V=V_0\e^{-\frac{2\phi}{\phi_0}}\ ,\quad
\xi_1=\xi_0\e^{\frac{2\phi}{\phi_0}}\ ,\quad
\xi_2=\eta_0\e^{\frac{4\phi}{\phi_0}}\ .
\ee
Here $\phi_0$, $V_0$, $\xi_0$, and $\eta_0$ are constants.
We may consider the limit of $\epsilon\to 0$ after redefining
$\phi=\epsilon\varphi$ and
$\phi_0=\epsilon\varphi_0$. Thus, Eq.(\ref{Eb4}) gives
\be
\label{E8}
\e^{\frac{2\varphi}{\varphi_0}} = \Theta (G, {\cal L}_c^{(2)})
\equiv \frac{\xi_0 G}{2\eta_0 {\cal L}_c^{(2)}} + Y(G, {\cal L}_c^{(2)})\ .
\ee
Here
\be
\label{E8b}
Y(G,{\cal L}_c^{(2)}) = y_+ + y_ - \ ,\quad
y_+ \e^{\frac{2}{3}\pi i} + y_- \e^{\frac{4}{3}\pi i} \ ,\quad y_+
\e^{\frac{4}{3}\pi i} + y_- \e^{\frac{2}{3}\pi i}
\ee
and
\be
\label{E8c}
y_\pm \equiv \left\{\frac{V_0}{4\eta_0 {\cal L}_c^{(2)}} \pm
\sqrt{\left(\frac{V_0}{4\eta_0 {\cal L}_c^{(2)}}\right)^2 - \left(\frac{\xi_0
G}{6\eta_0
{\cal L}_c^{(2)}}\right)^6}
\right\}^{1/3}\ .
\ee
Hence, the action of the corresponding $F(G,{\cal L}_c^{(2)})$-theory is
\bea
\label{E9}
S&=&\int d^4 x \sqrt{-g}\left[ \frac{R}{2\kappa^2} - F(G,{\cal
L}_c^{(2)})\right]\ ,\nn
F(G,{\cal L}_c^{(2)})&= & \frac{V_0}{\Theta\left(G,{\cal L}_c^{(2)}\right)} -
\xi_0\Theta\left(G,{\cal L}_c^{(2)}\right) G
+ \eta_0\Theta\left(G,{\cal L}_c^{(2)}\right)^2{\cal L}_c^{(2)}\ .
\eea

Instead of (\ref{GBany1}), one may consider the model with one more scalar
field
$\chi$
coupled with the Gauss-Bonnet
invariant:
\be
\label{GBany51}
S=\int d^4 x \sqrt{-g}\left[ \frac{R}{2\kappa^2} - \frac{1}{2}\partial_\mu \phi
\partial^\mu \phi
    - \frac{\epsilon}{2}\partial_\mu \chi \partial^\mu \chi
    - V(\phi) - U(\chi) - \left(\xi_1(\phi) + \theta(\chi)\right)G\right]\ .
\ee
This kind of action  often appears in the models inspired by the
string
theory \cite{ART}.
In such models, one scalar $\phi$ may correspond to the dilaton and another
scalar $\chi$ to modulus.
We now consider the case that the derivative of $\chi$, $\partial_\mu \chi$, is
small or $\epsilon$ is very small.
Then we may neglect the kinetic term of $\chi$ and $\chi$ could be regarded as
an auxiliary field. Repeating
the process (\ref{GBany44}-\ref{GBany46}), we obtain the $F(G)$-gravity coupled
with the scalar field $\phi$:
\be
\label{GBany52}
S=\int d^4 x \sqrt{-g}\left[ \frac{R}{2\kappa^2} - \frac{1}{2}\partial_\mu \phi
\partial^\mu \phi
    - V(\phi) - \xi_1(\phi) G + F(G)\right]\ .
\ee

The relation between scalar-Gauss-Bonnet gravity and modified Gauss-Bonnet
gravity (or two parameterizations of the same theory) is discussed in this
section. It is shown that cosmological solutions
obtained in one of such theories may be used (with different physical
interpretation , compare with \cite{salvatore}) in another theory.
It is often turns out that it is easier to work with specific
parametrization of the same theory. Of course, only comparison with
observational data may select the truly physical theory in correct
parametrization.

\section{Conclusion}

In this paper we have studied several aspects of (dilaton) gravity in the
presence of string corrections up to
third order in curvature. The second order term is Euler density of order two
called, the Gauss-Bonnet term.
The next-to-leading  term contains higher order Euler density ($E_3$)
plus a term of order three
in curvature. The expression of $E_3$ is identically zero in space-time of
dimension less than six; the term beyond the Euler
density contributes
to equation of motion even for a fixed field $\phi$. We have verified that the
de-Sitter solution which
exists in the case of Type II and Bosonic strings is an unstable node. It
is shown that in the presence
of a barotropic fluid (radiation/matter), inflationary solution exists in the
high curvature regime for constant field.

For a dynamically evolving field $\phi$  canonical in nature, there exists an
interesting
dark energy solution (\ref{solution}) characterized by $H={h_0}/{t}$, $\phi=\phi_0\ln{t}/{t_1}$
for $ h_0>0$
($H={h_0}/{t_s-t},\ \phi=\phi_0\ln(t_s-t/{t_1})\ (\mbox{when}\ h_0<0 $). The
three years
WMAP data taken with the SNL survey\cite{Spergel} suggests that
$w_{DE}=-1.06^{+0.13}_{-0.08}$. We have shown that
choosing a range of parameter $\phi_0^2 $ (which is amplified thanks to third
order curvature term contribution) we can easily obtain the observed
values of  $w_{DE}$
for phantom as well as for non-phantom dark energy.
We have demonstrated, in detail, the stability of dark energy solution. For non-phantom energy, 
in the large $h_0$
limit, we presented analytical solution which shows that one of the eigenvalues of the $3 \times 3$
perturbation matrix is real and negative  where as the other two are purely imaginary, thereby,
establishing the stability of solution (\ref{solution}). We have verified numerically that stability holds
for all smaller and generic values of $h_0$ in this case.  The phantom dark energy solution 
corresponding to $h_0<0$ turns out to be unstable.
 It is remarkable that
string curvature corrections
can account for late time acceleration and dark energy can be realized
without the introduction of a field potential.

It is shown how scalar-Gauss-Bonnet gravity may be reconstructed for any
given cosmology. The corresponding scalar potentials for several
dark energy cosmologies including quintessence, phantom, cosmological
constant or oscillatory regimes are explicitly found.
This shows that having the realistic scale factor evolution, the principal
possibility appears to present string-inspired gravity where such
evolution is realized. It is explained how to transform
scalar-Gauss-Bonnet gravity (even with account of third order curvature term)
to modified Gauss-Bonnet gravity \cite{GB} which seems to pass
the Solar System tests.

Different forms of modified gravity are attempted recently
(for a review, see \cite{rev3}) to describe dark energy universe;
these models provide a qualitatively simple resolution of dark
energy/coincidence problems and deserve further consideration.
It is quite likely that time has come to reconsider the basics of General
Relativity at the late universe in the search of realistic modified
gravity/dark energy theory.

We should also mention that in the present study we have tested the 
background model against observations. The study of perturbations 
in the scenario discussed here is quite complicated and challenging
and in our opinion it deserves attention; we defer this investigation to our future work.
 
\section*{Acknowledgments}

The research of SDO is supported in part by the project
FIS2005-01181 (MEC, Spain), by LRSS project N4489.2006.02 and by
RFBR grant 06-01-00609 (Russia). MS thanks S. Panda, I.~Neupane  and S. Tsujikawa and SDO
thanks M. Sasaki for
useful discussions.

\end{document}